\documentclass[twocolumn,prl]{revtex4-1}
\pdfoutput=1

\usepackage{amsfonts}
\usepackage{amsmath}
\usepackage{amssymb}
\usepackage{color}
\usepackage{bm}
\usepackage{graphicx}
\usepackage{natbib}
\usepackage{hyperref}
\hypersetup{
	colorlinks = true
}
\usepackage{natbib}

\begin{document}
	\title{Josephson radiation in a superconductor-quantum dot-superconductor junction}
	\author{Baptiste Lamic}
	\author{Julia S. Meyer}
	\author{Manuel Houzet}
	\affiliation{Univ.~Grenoble  Alpes,  CEA,  IRIG-Pheliqs,  F-38000  Grenoble,  France}
	\date{\today}
	\begin{abstract}
		We investigate the Josephson radiation emitted by a junction made of a quantum dot coupled to two conventional superconductors. Close to resonance, the particle-hole symmetric Andreev states that form in the junction are detached from the continuum above the superconducting gap in the leads, while a gap between them opens near the Fermi level. Under voltage bias, we formulate a stochastic model that accounts for non-adiabatic processes, which change the occupations of the Andreev states. This model allows calculating the current noise spectrum and determining the Fano factor. Analyzing the finite-frequency noise, we find that the model may exhibit either an integer or a fractional AC Josephson effect, depending on the bias voltage and the size of the gaps in the Andreev spectrum. Our results assess the limitations in using the fractional Josephson radiation as a probe of topology.
	\end{abstract}
	\maketitle
	
	The Josephson radiation is the electromagnetic signal emitted by a junction between two superconductors when it is voltage biased. Its measurement is the first experimental demonstration of the AC Josephson effect~\cite{Yanson1965}. The charge of the Cooper pairs, which form the superconducting condensate, appears in the relation between the radiation frequency and voltage bias, $\omega=\omega_J$, where $\omega_J= 2eV/\hbar$ is the Josephson frequency. The coherence property of the Josephson radiation is usually limited by the electromagnetic environment of the junction. Tailoring the environment surrounding a superconducting tunnel junction recently allowed reaching the regime of Josephson lasing~\cite{Cassidy2017}. 
	
	The interest in the Josephson radiation was revived by the prediction of a fractional AC Josephson effect at frequency $\omega=\omega_J/2$ when the superconducting leads forming the junction are topological~\cite{Kitaev2001,Kwon2004,Deacon2017JosephsonJunctions,Laroche2019ObservationNanowires}. This fractional Josephson radiation originates from the fact that the supercurrent flowing through the junction is carried by single electrons, rather than Cooper pairs. A topological junction admits two degenerate parity sectors, carrying Josephson supercurrents with opposite values. Therefore, random parity switchings generate current noise, and eventually bring another limitation to the coherence of the Josephson radiation in the topological case~\cite{Fu2009}.
	
	On a microscopic level, the Josephson effect can be associated with the formation of subgap states, also known as Andreev bound states (ABS), in the junction. The difference between conventional and topological Josephson junctions is associated with the fact that, in equilibrium, the ABS energy depends $2\pi$-periodically on the superconducting phase difference $\varphi$ in the conventional case, while the dependence is $4\pi$-periodic in the topological case. Then the integer or fractional Josephson radiations simply result from the substitution $\dot\varphi=\omega_J$ in the phase dependence of the current carried by an ABS whose occupation is fixed. However, such considerations neglect  the voltage-induced non-adiabatic processes that change the ABS occupations and, therefore, reduce the coherence of the Josephson radiation, even when the effect of the external environment is negligeable. Indeed, it was shown that such non-adiabatic processes in topological junctions introduce an intrinsic limitation to the visibility of the fractional Josephson radiation~\cite{Badiane2011,Pikulin2012,San-Jose2012}.  This mechanism is ultimately related with the dissipative current that flows through the junction, which is induced by these processes.
	
	In this work, we investigate the role of non-adiabatic processes on the Josephson radiation in a {\it non-topological} junction formed by a quantum dot connected to conventional superconducting leads. We find that such conventional junctions may display either a conventional or a fractional Josephson radiation as the voltage varies, depending on details of the ABS spectrum. In particular, a fractional Josephson radiation is predicted when the gap in the Andreev spectrum near the Fermi level is crossed diabatically, while the ABS are sufficiently detached from the continuum above the superconducting gap. Furthermore, we determine the contribution of the non-adiabatic transitions to the linewidth of the Josephson radiation. Our results assess the limitations in using the current noise spectrum to determine whether a Josephson junction is topological or not.

	\vspace{0.5cm}

\begin{figure}
		\centering
		\includegraphics[width=\linewidth]{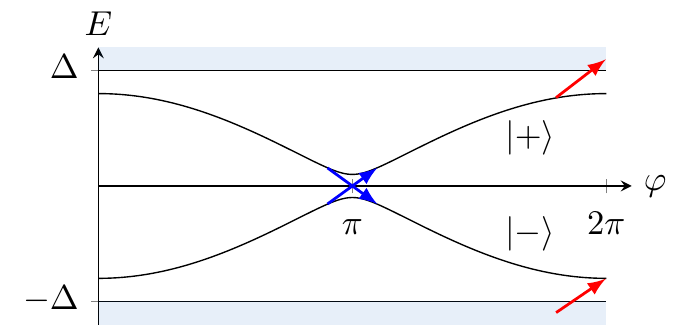}
		\caption{
			\label{fig:spectrum}
			Spectrum of the Andreev bound states as a function of the superconducting phase difference. In the presence of a voltage bias, the phase increases with time and non-adiabatic transitions may occur between states $|+\rangle$ and $|-\rangle$ (blue arrows),
			as well as {between} state $|+\rangle$ and the continuum at energy $E>\Delta$ {or between the continuum at energy $E<-\Delta$ and state $|-\rangle$}  (red arrows).
		}
	\end{figure}

	We consider a junction made of a spin-degenerate single-level quantum dot that is contacted to two superconducting leads. We first review the properties of its Andreev spectrum, which have been analyzed both in the presence and absence of Coulomb interaction \cite{Martin-Rodero2011}. Let us first start with the case of strong Coulomb interaction. It allows for resonant scattering of electrons between the leads in the normal state thanks to the Kondo effect. For energies smaller than the Kondo scale $T_K$, the transmission probability is $T_\pi=4\Gamma_L\Gamma_R/\Gamma^2$, where $\Gamma=\Gamma_L+\Gamma_R$ and $\Gamma_L$ and $\Gamma_R$ are the partial level widths due to the coupling between the dot and the left and right leads. For an almost symmetric coupling, the transmission is almost ballistic, $R_\pi\equiv 1-T_\pi\ll 1$. When the leads are superconducting and the superconducting gap is sufficiently small, $\Delta\ll T_K$, the junction accommodates two particle-hole symmetric ABS denoted $|+\rangle$ and $|-\rangle$. Their energies are approximated by   
	\begin{equation}
	\label{eq:ABS}
	E_\pm(\varphi)=\pm\Delta\sqrt{T_0\cos^2(\varphi /2)+R_\pi\sin^2(\varphi /2)}.
	\end{equation}
	In addition to the gap opening in the Andreev spectrum near $\varphi=\pi$ at $T_\pi<1$~\cite{Glazman1989ResonantBarrier}, Eq.~\eqref{eq:ABS} also shows the detachment of the ABS from the edge of the continuum spectrum near $\varphi=0$~\cite{Vecino2003,Zazunov2018}, which is controlled by an effective reflection probability $R_0\equiv 1-T_0$ of order $(\Delta/T_K)^2\ll 1$ (up to a logarithmic correction~\cite{Matsuura1977}). Interestingly, Eq.~\eqref{eq:ABS} also holds in the absence of interactions, after substituting $T_K$ with $\Gamma\gg\Delta$~\cite{Beenakker1991}. The ABS dispersion is illustrated in Fig.~\ref{fig:spectrum}.
	
	We turn now to the current flowing through the junction. Each occupied ABS carries a supercurrent
	\begin{equation}
	I_\pm(\varphi)=\frac{2e}\hbar \frac{\partial E_\pm(\varphi)}{\partial\varphi}
	\approx \mp I_J\sin\frac\varphi 2\text{sign}\left(\cos\frac\varphi 2\right),
	\end{equation}
	where $I_J=e\Delta/\hbar$, and we used $R_0,R_\pi\ll 1$ in the last equality. In equilibrium, the average ABS occupations are set by the Fermi distribution, while the contribution of the continuum is negligibly small. 
	Thus the equilibrium supercurrent at zero temperature is given by $I_-(\varphi)$.
	
	In the presence of a dc voltage bias, the phase difference increases linearly with time, $\varphi(t)=\omega_J t+\varphi_0$ with a reference phase $\varphi_0$. (Here we assume $V>0$,  for concreteness.) As a consequence, changes in the occupations of the ABS can occur due to non-adiabatic transitions. Using $R_0,R_\pi \ll 1$ and $V\ll \Delta/e$, we can isolate two kinds of non-adiabatic processes. Near $\varphi=\pi\,\text{mod}\, 2\pi$, these are the transitions between $|+\rangle$ and $|-\rangle$, which occur with {the} Landau-Zener probability $p_\pi=\exp(-\pi R_\pi\Delta/eV)$~\cite{Averin1995}; $p_\pi$ increases rapidly from 0 to 1 as $V$ increases in the vicinity of $V_\pi=R_\pi\Delta/e$. Near $\varphi=0\,\text{mod}\, 2\pi$, these are non-adiabatic transitions between $|+\rangle$ and the continuum states with energy $E>\Delta$ as well as the continuum states with energy $E<-\Delta$ and $|-\rangle$, which take place with probability $p_0=p(V/V_0)$ with $V_0=R_0^{3/2}\Delta/e$, 
	where the function $p(x)$ calculated in Ref.~\cite{Houzet2013} (and Ref.~\cite{Yeyati2003} at $V\ll V_0$) is such that $p_0$ increases rapidly from 0 to 1 as $V$ increases in the vicinity of $V_0$.

\begin{figure}
		\centering
		\includegraphics[width=\linewidth]{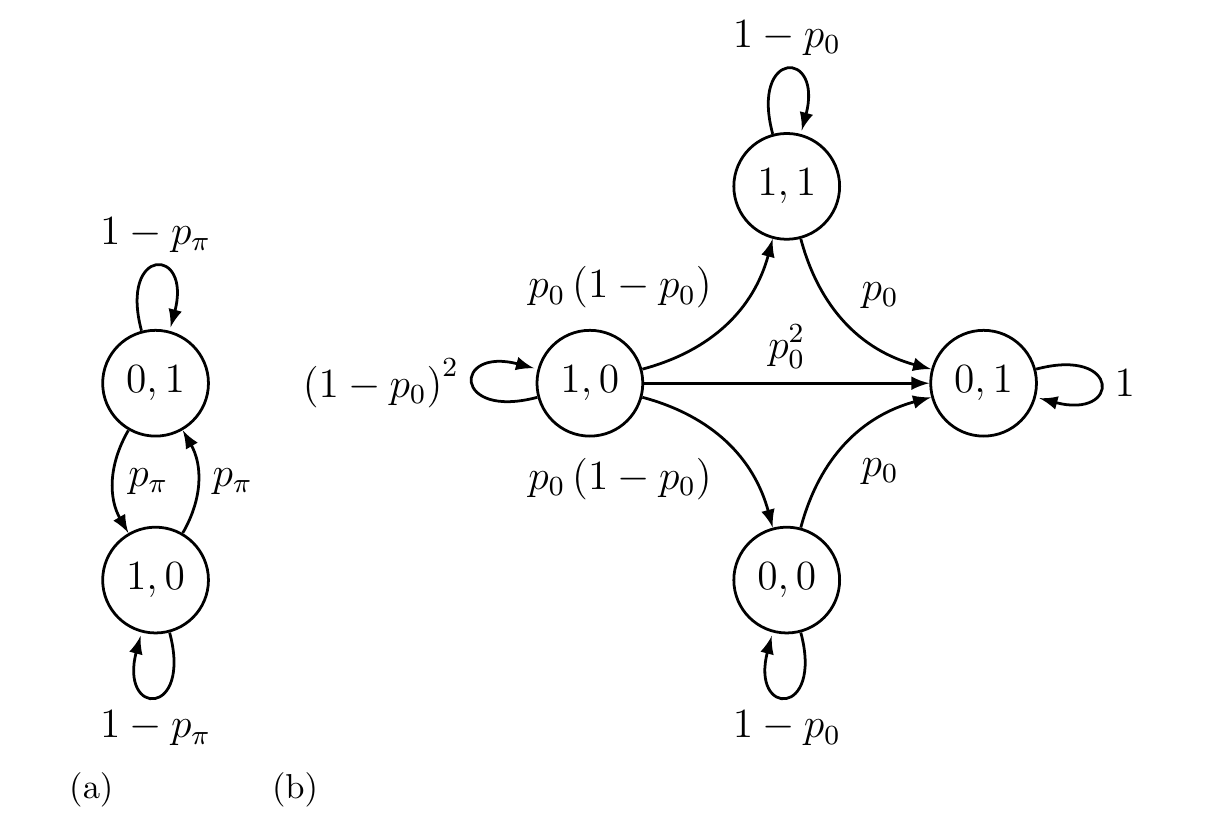}
		\caption{
			\label{fig:L}
		 Graph of the transition matrices. Each arrow denotes a possible transition between two states $(n_+,n_-)$ and $(n'_+,n'_-)$, and is labelled by the transition probability. (a) Non adiabatic processes described by ${\cal\bm{L}}_\pi$ can exchange the occupation of states $(0,1)$ and $(1,0)$ by enabling a particle from one ABS to jump {to} the other ABS. (b) Non adiabatic processes described by ${\cal\bm{L}}_0$ can populate the lower ABS, enabling transitions from  $(0,0)$ to $(0,1)$, and from $(1,0)$ to $(1,1)$. They can also deplete the higher ABS state, enabling transitions from $(1,0)$ to $(0,0)$ and from $(1,1)$ to $(0,1)$.}
	\end{figure}

	We assume $T\ll \Delta$, so that continuum states with energy $E<\Delta$ ($E>\Delta$) are occupied (empty). We also neglect the short timescales over which the non-adiabatic processes take place on the scale of the Josephson period, $2\pi/\omega_J$. 
	Then at each time, the state of the junction is fully characterized by the occupations $n_\pm=0,1$. (In particular, we ignore coherent superpositions between $|\pm\rangle$-states.) The states $(0,1)$ and $(1,0)$ are the ground and first excited states in the even parity sector of the junction, respectively; the states $(0,0)$ and $(1,1)$ are the ``poisoned'' states in the odd parity sector. Within a Markov model that describes switches in their random occupations~\cite{Averin1996,Houzet2013}, the average supercurrent is
	\begin{equation}
	\label{eq:I}
	\langle I(t)\rangle =
	{\cal I}(t)\left[P_{01}(\varphi(t))-P_{10}(\varphi(t))\right].
	\end{equation}
	Here ${\cal I}(t)=I_-\left(\varphi(t)\right)$ and $P_{n_+n_-}(\varphi)$ with $n_+,n_-=0,1$ {denotes} the ABS occupations at a given phase. Neglecting any coupling with an external bath, these probabilities remain constant within intervals $\pi m<\varphi<\pi (m+1)$ with integer $m$, while their values immediately before and after the specific phases where non-adiabatic transitions can take place are related with each other {through} the transition probabilities $p_0$ and $p_\pi$,
	\begin{subequations}
		\label{eq:stoch}
		\begin{eqnarray}
		\bm{P}(2m\pi^+)&=&{\cal\bm{L}}_0\bm{P}(2m\pi ^-),
		\\
		\bm{P}((2m+1)\pi^+)&=&{\cal\bm{L}}_\pi\bm{P}((2m+1)\pi ^-)
		\end{eqnarray}
	\end{subequations}
	with $\bm{P}(\varphi)=\left(P_{11}(\varphi),P_{10}(\varphi),P_{01}(\varphi),P_{00}(\varphi)\right)^T$ and the transition matrices
	\begin{subequations}
		\label{eq:mat-stoch}
		\begin{eqnarray}
		{\cal\bm{L}}_0 &=& 
		\left(\begin{array}{cccc}
		1-p_0& p_0(1-p_0) & 0 & 0\\
		0 & (1-p_0)^2 & 0 & 0\\
		p_0 & p_0^2 & 1 & p_0\\
		0 & p_0(1-p_0) & 0 & 1 - p_0
		\end{array}\right),
		\\
		{\cal\bm{L}}_\pi &=&  
		\left(\begin{array}{cccc}
		1&0&0&0\\
		0&1- p_\pi &p_\pi&0
		\\0&p_\pi&1- p_\pi&0
		\\0&0&0&1
		\end{array}\right).
		\label{eq:Lpi}
		\end{eqnarray}
	\end{subequations}
	The non-adiabatic processes accounted {for} by the matrix elements of ${\cal\bm{L}}_0$ and ${\cal\bm{L}}_\pi$ are illustrated in Fig.~\ref{fig:L}.
	
	The stochastic matrix ${\cal\bm{L}}\equiv{\cal\bm{L}}_0{\cal\bm{L}}_\pi$ admits normalized right- and left-eigenvectors $u_\alpha$ and $v_\alpha$, with a common eigenvalue $\lambda_\alpha$, such that ${\cal\bm{L}}u_\alpha=\lambda_\alpha u_\alpha$,  ${\cal\bm{L}}^Tv_\alpha=\lambda_\alpha v_\alpha$, and $v^T_\alpha u_\beta=\delta_{\alpha\beta}$. Furthermore, the eigenvalue $\lambda_0=1$ is associated with the left eigenvector $v_0=(1,\dots,1)^T$, while other eigenvalues,
	$\lambda_1=1-p_0$, $\lambda_2=(1-p_0)^2$, and $\lambda_3=(1-p_0)(1-2p_\pi)$, satisfy $|\lambda_{\alpha\neq 0}|<1$. Thus, the probability vector $\bm P$ that solves Eqs.~\eqref{eq:stoch} reaches a solution at long times that does not depend on {the} initial condition; it is {given by} the eigensolution $\alpha=0$:
	\begin{equation} 
	\label{eq:P}
	\bm{P}(\varphi)=\left\{\begin{array}{ll}
	u_0, & 2 m\pi<\varphi<(2m+1)\pi,\\
	{\cal\bm{L}}_\pi u_0, & (2 m+1)\pi<\varphi<(2m+2)\pi,
	\end{array}
	\right.
	\end{equation}
	with 
	\begin{subequations} 
		\begin{eqnarray} 
		u_0&=&\frac 1 {\cal N}
		\left(\begin{array}{c}
		p_\pi(1-p_0)\\
		p_\pi(1-p_0)^2\\
		{\cal N}-p_\pi(1-p_0)(3-p_0)\\
		p_\pi(1-p_0)
		\end{array}
		\right),
		\\
		{\cal\bm{L}}_\pi u_0&=&\frac 1 {\cal N}\left(\begin{array}{c}
		p_\pi(1-p_0)\\
		p_\pi\\
		{\cal N}-p_\pi(3-2p_0)\\
		p_\pi(1-p_0)
		\end{array}
		\right),
		\end{eqnarray}
		\label{eq:u0}
	\end{subequations}
	and ${\cal N} = (2-p_0)(1-\lambda_3)$. In particular, the ground state $(0,1)$ is mostly occupied with $u_0\approx  {\cal\bm{L}}_\pi u_0 \approx (0,0,1,0)^T$ at $p_\pi\ll p_0$, while all states are approximately equally occupied with $u_0\approx  {\cal\bm{L}}_\pi u_0 \approx (\frac 14,\frac 14,\frac 14,\frac 14)^T$ at $p_0\ll p_\pi$.
	
	Inserting Eq.~\eqref{eq:P} into \eqref{eq:I}, we find the average current in the long-time limit,
	\begin{eqnarray}
	\label{eq:I2}
	\langle I(t)\rangle&=&\frac{p_0}{1-\lambda_3}\left[p_\pi \left|{\cal I}(t)\right|+(1-p_\pi) {\cal I}(t) \right]
	\\&=&
	I_{\text{dc}}\Bigg\{ 1 -\sum_{n\geq 1}\frac 2{4n^2-1}\Big[ \cos(n\varphi(t))
	\nonumber\\&&
	\qquad\qquad+2(-1)^nn\frac{1-p_\pi}{p_\pi} \sin(n\varphi(t)) \Big]\Bigg\}.\;\;
	\nonumber
	\end{eqnarray}
	The dc contribution, 
	\begin{equation}
	\label{eq:Idc}
	I_{\text{dc}}=  \frac 2\pi \frac{p_0 p_\pi}{1-\lambda_3}I_J,
	\end{equation}
	relates the dissipative current with non-adiabatic processes through the gaps in the Andreev spectrum. It corresponds to the low-voltage regime of multiple Andreev reflections (MAR). It generalizes formulas derived in superconducting atomic contacts~\cite{Averin1995} at $p_0=1$, and in topological Josephson junctions~\cite{Houzet2013} at $p_\pi=1$. The ratio between {cosine and sine} harmonics of the Josephson frequency is controlled by non-adiabatic transitions between ABS with positive and negative energies. When these processes are rare, the {sine} harmonics dominate like in the adiabatic case.

	Due to the stochastic nature of the non-adiabatic processes, the current fluctuates. We characterize the fluctuations with the current noise spectrum, 
	\begin{equation}
	\label{eq:defnoise}
	S(\omega)=2\int_0^\infty d\tau  \cos(\omega\tau) \overline{S(t+\tau/2,t-\tau/2)},
	\end{equation}
	where the bar denotes an average over $t$. Within the Markov theory, we relate the current correlator,
	\begin{equation}
	\label{eq:S}
	S(t_1,t_2)=\langle I(t_1)I(t_2)\rangle-\langle I(t_1)\rangle \langle I (t_2)\rangle,
	\end{equation}
	with
	\begin{eqnarray}
	\label{eq:corr-curr}
	&&\langle I(t_1)I(t_2)\rangle = {\cal I}(t_1){\cal I}(t_2) 
	\\
	&&\qquad\qquad
	\times \sum_{n_1,n_2
	}(-1)^{n_1+n_2} P_{n_1\bar n_1|n_2\bar n_2}(\varphi_1|\varphi_2) 
	P^\infty_{n_2\bar n_2}(\varphi_2)
	\nonumber
	\end{eqnarray}
	at $t_1>t_2$. Here $P_{n_1 n'_1|n_2 n'_2}(\varphi_1|\varphi_2)$ with $\varphi_i = \varphi(t_i)$ is the conditional probability for the system to reside in state $(n_1,n'_1)$ at phase $\varphi_1$ if it was in state $(n_2,n'_2)$ at phase $\varphi_2<\varphi_1$; it solves the same Eq.~\eqref{eq:stoch} as the probability $P_{n_1n_1'}(\varphi_1)$, together with the initial condition $P_{n_1n_1'|n_2n'_2}(\varphi_2|\varphi_2)=\delta_{n_1,n_2}\delta_{n'_1,n'_2}$. Furthermore we used notations $\bar0=1$ and $\bar 1=0$. Using a matrix representation [in the same basis of states as the one used in Eqs.~\eqref{eq:stoch} and \eqref{eq:mat-stoch}] for the closure relation, $\sum_\alpha u_\alpha v^T_\alpha=1$, we find
	\begin{equation}
	\label{eq:cond-prob}
	P(\varphi_1|\varphi_2)=
	{\cal \bm{L}}_\pi^{k_1}\left(\sum_\alpha u_\alpha\lambda_\alpha^{m_1-m_2}v_\alpha^T\right) {\cal \bm{L}}_\pi^{-k_2}
	\end{equation}
	for  $(2 m_i+k_i)\pi<\varphi_i<(2 m_i+k_i+1)\pi$ with $m_i$ integer, $k_i=0,1$, and $i=1,2$. The second term in the r.h.s.~of Eq.~\eqref{eq:S} compensates the contribution from the terms with $\alpha=0$ when inserting Eq.~\eqref{eq:cond-prob} into \eqref{eq:corr-curr}. Furthermore, the expectation value of the current operator {in} the states with $\alpha=1,2$ vanishes. Therefore, only the terms with $\alpha=3$ contribute to Eq.~\eqref{eq:S}. As the result of a long calculation described in Sec.~S1 of the supplemental material (SM)~\cite{supplemental}, we find  
	\begin{equation}
	\label{eq:spectrum}
	\frac{S(\omega)}{ S_0}=
	\frac{
		A \left[1+4 \tilde\omega ^2-4 \tilde\omega  \sin (\pi  \tilde\omega )\right]+(B + C  \tilde\omega ^2) \cos ^2(\pi  \tilde\omega)}
	{  (1-4\tilde\omega^2)^2 \left[\left(1+\lambda_3\right)^2-4\lambda_3\cos^2(\pi \tilde\omega)\right]}
	\end{equation}
	with $S_0={ I_J^2  }/({\pi  \omega_J})$, $\tilde\omega=\omega/\omega_J$, and
	\begin{subequations}
		\begin{eqnarray}
		A&=&8p_0p_\pi(1-p_\pi)[1+(1-p_0)^2](1+\lambda_3)/{\cal N},\\
		B&=&-16p_\pi(2-p_\pi)\lambda_3/{\cal N},\\
		C&=&128p_\pi(1-p_\pi)(1-p_0)[1-\lambda_3(1-p_0)]/{\cal N}.\qquad
		\end{eqnarray}
	\end{subequations}
	Below we {discuss the zero-frequency noise as well as structures related to the AC currents}  in the frequency dependence of the noise given by Eq.~\eqref{eq:spectrum}.
	
\begin{figure}
		\centering
		\includegraphics[width = \linewidth]{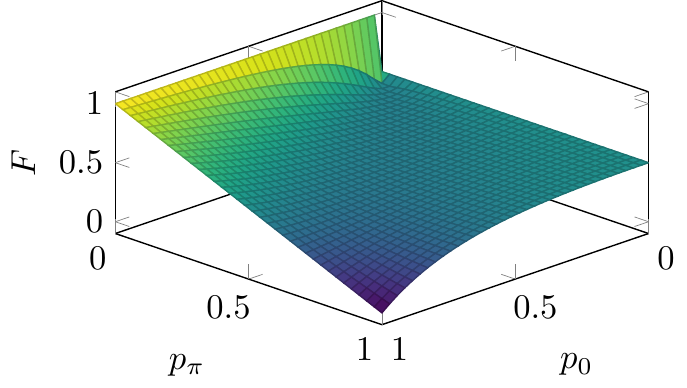}
		\caption{
			\label{F:Fano} 
			Fano factor as a function of $p_0$ and $p_\pi$.
		}
	\end{figure}

	The zero-frequency noise is expressed in terms of an effective charge, $e^\star=2\Delta/V$, which diverges inversely with the bias voltage in the MAR regime, and a Fano factor
	\begin{eqnarray}
	\label{eq:fano}
	F\equiv\frac{S(0)}{e^\star I_{\text{dc}}}=\frac{A+B}{8p_0p_\pi(1-\lambda_3)}.
	\end{eqnarray}
	In particular, $F=1$ at $p_\pi\ll p_0$, when the bottleneck for the transfer of quasiparticles across the junction is the gap near the Fermi level~\cite{Naveh1999}; $F=1/2$ at $p_0\ll p_\pi$, when the bottleneck consists of the two gaps {(of the same width)} near the edges of the continuum spectrum. In particular, at $p_\pi=1$, on recovers the results of Ref.~\cite{Houzet2013} for a topological junction (see Sec.~S2 of the SM~\cite{supplemental}). $F$ vanishes at $ 1-p_0, 1-p_\pi\ll 1$, when quasiparticle transfer across the gap {becomes} deterministic. In the general case $0<F<1$, see Fig.~\ref{F:Fano}.
	
	{From Eq.~\eqref{eq:spectrum}, we see that the noise displays sharp features when $1\mp\lambda_3\ll1$. This is the case when $p_0,p_\pi\ll 1$ or $p_0,1-p_\pi\ll 1$.
		We turn now to the noise spectrum in these limiting cases for the transition probabilities.} 
	
	At $p_0,p_\pi\ll 1$ the noise spectrum displays features with a narrow linewidth,  \begin{equation}
	\gamma=\frac 1{2\pi}(p_0+2p_\pi){\omega_J},
	\end{equation}
	near $\omega={n}\omega_J$ with ${n}$ integer. In particular, at very low frequency, $\omega\ll \omega_J$, 
	\begin{equation}
	\label{eq:Feff}
	\frac{S(\omega)}{e^\star  I_{\text{dc}}}=F+(F_{\text{app}}-F)\frac{\omega^2}{\omega^2+\gamma^2}.
	\end{equation}
	Here, Eq.~\eqref{eq:Idc} simplifies to
	\begin{equation}
	I_{\text{dc}}=\frac {2I_J}\pi\frac{p_0p_\pi}{p_0+2p_\pi},
	\end{equation} and the Fano factor,
	\begin{equation}
	F=\frac{p_0^2+p_0p_\pi+2p_\pi^2}{(p_0+2p_\pi)^2},
	\end{equation}
	is only accessible in a narrow frequency range {$\omega\ll\gamma$}, while an apparent Fano factor,
	\begin{equation}
	F_{\text{app}}=\frac{(12-4\pi+\pi^2)p_0+8p_\pi}{2 \pi^2 p_0},
	\end{equation}
	characterizes the noise in a wide frequency range $\gamma\ll\omega\ll \omega_J$. Note that $F_{\text{app}}\gg F$ if $p_0\ll p_\pi$.
	Furthermore 
	\begin{equation}
	\label{eq:noise-smallp}
	{S(\omega)}=\frac{32n^2}{(4n^2-1)^2\pi}\frac{(3p_0+2 p_\pi)p_\pi}{(p_0+2p_\pi)^2}\frac{\gamma I_J^2/\pi}{(\omega-{n}\omega_J)^2+\gamma^2}
	\end{equation}
	at $|\omega-{n} \omega_J|\ll \omega_J$, up to a negative resonance-frequency shift of the order of $\gamma^2/\omega_J\ll\gamma$. Comparing Eq.~\eqref{eq:noise-smallp} with the amplitude of the harmonics in Eq.~\eqref{eq:I2}, we conclude that the Josephson radiation at $p_0\ll p_\pi\ll 1$ is dominated by the noise, Eq.~\eqref{eq:noise-smallp}; thus it is  broadened by non-adiabatic transitions. On the other hand, the Josephson radiation at $p_\pi\ll p_0\ll 1$ is dominated by the {sine} harmonics in Eq.~\eqref{eq:I2}; thus it is broadened by the environment of the junction.
	
	At $p_0,1-p_\pi\ll 1$,  $I_{\text{dc}}=p_0 I_J /\pi$ and $F=1/2$;
	the noise spectrum displays a narrow resonance at half the Josephson frequency
	\begin{equation}
	S(\omega)=\frac{I_J^2}{4}\frac{\gamma'}{(\omega-\omega_J/2)^2+\gamma'^2}
	\end{equation}
	with linewidth 
	\begin{equation}
	\label{eq:noise-smallp2}
	\gamma'=\frac{1}{2\pi}{[p_0+2(1-p_\pi)]\omega_J},
	\end{equation} 
	up to a small resonance-frequency shift, of the order of $\gamma'^2/\omega_J\ll\gamma'$, which increases as $p_\pi$ decreases. Higher order resonances around $({n}+1/2)\omega_J$ are suppressed. Comparison between Eqs.~\eqref{eq:I2} and \eqref{eq:noise-smallp2} shows that the Josephson radiation is dominated by the noise, and it is thus broadened by the non-adiabatic processes. At $1-p_\pi \ll1$, transitions across the gap at $\pi$ are very frequent, leading to a large, but random $4\pi$-periodic contribution to the current. Thus, the Josephson radiation is fractional despite the junction {\it not} {being} topological. In the extreme case $p_\pi=1$, where the transitions across the gap at $\pi$ are deterministic, the system can be described as two independent topological junctions in parallel~\cite{supplemental,Zazunov2018}.

\begin{figure}
		\centering
		\includegraphics[width=\linewidth]{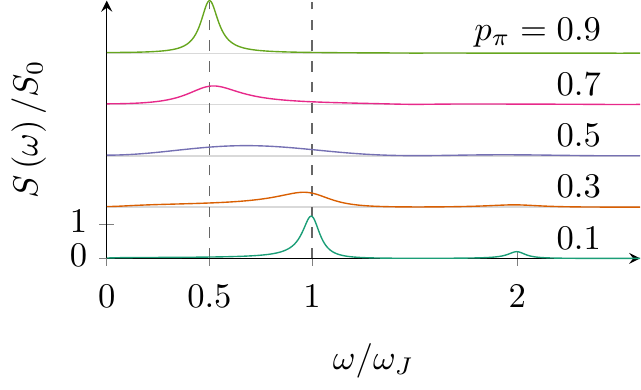}
		\caption{\label{fig:transition}
			Current noise spectrum for $p_0 = 0.1$ and several equally spaced values of $p_\pi$ between $0.1$ and $0.9$. The curves are shifted vertically for clarity.
		}
	\end{figure}

	The crossover between a well-resolved fractional or conventional Josephson radiation, when $p_0$ is small and $p_\pi$ increases from 0 to 1 occurs through the gradual shift and broadening of the peaks in the noise spectrum, as illustrated in Fig.~\ref{fig:transition}. A similar behavior has been reported in topological junctions, but with a residual coupling to the Majoranas at the far ends of the superconducting wires~\cite{Pikulin2012}.

	In a given junction, we can now distinguish two qualitatively different behaviors, depending on the ratio between crossover voltages $V_0$ and $V_\pi$.
	If $V_0\ll V_\pi$, $F=1$ at $V \ll V_\pi$, then it drops drops to 0 at $V \gg V_\pi$. Furthermore, the Josephson radiation is conventional. In the opposite case, $V_\pi \ll V_0$, $F=1/2$ at $V\ll V_0$, then it drops drops to 0 at $V \gg V_0$, and there is a voltage range $V_\pi \ll V \ll V_0$ where Josephson radiation is fractional. Both the conventional and fractional Josephson radiations are broadened by the non-adiabatic processes at $V<V_0$.
	
	\vspace{0.5cm}
	
	In conclusion we proposed a simple model to analyze the role of non-adiabatic transitions between Andreev states in the Josephson radiation of a superconductor-quantum dot-superconductor junction. Within a simplified model of the Andreev states' dynamics, we predicted that such a conventional junction may display either fractional or conventional radiation depending on its parameters and on the voltage bias. On the theory side, it would be interesting to extend the analysis to a more general description of the Andreev dynamics, as well as develop a theory frame to compare the interplay of the environment (neglected in our study) and the non-adiabatic transition on the loss {of  coherence} of the Josephson radiation. On the experimental side, it would be interesting to compare our prediction with finite-frequency noise {measurements} in devices such as the superconductor-carbon nanotube-superconductor junction in the Kondo region, whose dc transport and shot noise have been measured recently~\cite{Hata2018}.
	
	\acknowledgments{
		We acknowledge funding by the ANR through the grants ANR-16-CE30-0019 and ANR-17-PIRE-0001.}
	
	\bibliography{ms}

\begin{thebibliography}{23}%
\makeatletter
\providecommand \@ifxundefined [1]{%
 \@ifx{#1\undefined}
}%
\providecommand \@ifnum [1]{%
 \ifnum #1\expandafter \@firstoftwo
 \else \expandafter \@secondoftwo
 \fi
}%
\providecommand \@ifx [1]{%
 \ifx #1\expandafter \@firstoftwo
 \else \expandafter \@secondoftwo
 \fi
}%
\providecommand \natexlab [1]{#1}%
\providecommand \enquote  [1]{``#1''}%
\providecommand \bibnamefont  [1]{#1}%
\providecommand \bibfnamefont [1]{#1}%
\providecommand \citenamefont [1]{#1}%
\providecommand \href@noop [0]{\@secondoftwo}%
\providecommand \href [0]{\begingroup \@sanitize@url \@href}%
\providecommand \@href[1]{\@@startlink{#1}\@@href}%
\providecommand \@@href[1]{\endgroup#1\@@endlink}%
\providecommand \@sanitize@url [0]{\catcode `\\12\catcode `\$12\catcode
  `\&12\catcode `\#12\catcode `\^12\catcode `\_12\catcode `\%12\relax}%
\providecommand \@@startlink[1]{}%
\providecommand \@@endlink[0]{}%
\providecommand \url  [0]{\begingroup\@sanitize@url \@url }%
\providecommand \@url [1]{\endgroup\@href {#1}{\urlprefix }}%
\providecommand \urlprefix  [0]{URL }%
\providecommand \Eprint [0]{\href }%
\providecommand \doibase [0]{http://dx.doi.org/}%
\providecommand \selectlanguage [0]{\@gobble}%
\providecommand \bibinfo  [0]{\@secondoftwo}%
\providecommand \bibfield  [0]{\@secondoftwo}%
\providecommand \translation [1]{[#1]}%
\providecommand \BibitemOpen [0]{}%
\providecommand \bibitemStop [0]{}%
\providecommand \bibitemNoStop [0]{.\EOS\space}%
\providecommand \EOS [0]{\spacefactor3000\relax}%
\providecommand \BibitemShut  [1]{\csname bibitem#1\endcsname}%
\let\auto@bib@innerbib\@empty
\bibitem [{\citenamefont {Yanson}\ \emph {et~al.}(1965)\citenamefont {Yanson},
  \citenamefont {Svistunov},\ and\ \citenamefont {Dmitrenko}}]{Yanson1965}%
  \BibitemOpen
  \bibfield  {author} {\bibinfo {author} {\bibfnamefont {I.}~\bibnamefont
  {Yanson}}, \bibinfo {author} {\bibfnamefont {V.}~\bibnamefont {Svistunov}}, \
  and\ \bibinfo {author} {\bibfnamefont {I.}~\bibnamefont {Dmitrenko}},\ }\href
  {http://jetp.ac.ru/cgi-bin/e/index/e/21/3/p650?a=list
  http://adsabs.harvard.edu/abs/1965JETP...21..650Y} {\bibfield  {journal}
  {\bibinfo  {journal} {J. Exp. Theor. Phys.}\ }\textbf {\bibinfo {volume}
  {21}},\ \bibinfo {pages} {650} (\bibinfo {year} {1965})}\BibitemShut
  {NoStop}%
\bibitem [{\citenamefont {Cassidy}\ \emph {et~al.}(2017)\citenamefont
  {Cassidy}, \citenamefont {Bruno}, \citenamefont {Rubbert}, \citenamefont
  {Irfan}, \citenamefont {Kammhuber}, \citenamefont {Schouten}, \citenamefont
  {Akhmerov},\ and\ \citenamefont {Kouwenhoven}}]{Cassidy2017}%
  \BibitemOpen
  \bibfield  {author} {\bibinfo {author} {\bibfnamefont {M.~C.}\ \bibnamefont
  {Cassidy}}, \bibinfo {author} {\bibfnamefont {A.}~\bibnamefont {Bruno}},
  \bibinfo {author} {\bibfnamefont {S.}~\bibnamefont {Rubbert}}, \bibinfo
  {author} {\bibfnamefont {M.}~\bibnamefont {Irfan}}, \bibinfo {author}
  {\bibfnamefont {J.}~\bibnamefont {Kammhuber}}, \bibinfo {author}
  {\bibfnamefont {R.~N.}\ \bibnamefont {Schouten}}, \bibinfo {author}
  {\bibfnamefont {A.~R.}\ \bibnamefont {Akhmerov}}, \ and\ \bibinfo {author}
  {\bibfnamefont {L.~P.}\ \bibnamefont {Kouwenhoven}},\ }\href {\doibase
  10.1126/science.aah6640} {\bibfield  {journal} {\bibinfo  {journal}
  {Science}\ }\textbf {\bibinfo {volume} {355}},\ \bibinfo {pages} {939}
  (\bibinfo {year} {2017})}\BibitemShut {NoStop}%
\bibitem [{\citenamefont {Kitaev}(2001)}]{Kitaev2001}%
  \BibitemOpen
  \bibfield  {author} {\bibinfo {author} {\bibfnamefont {A.~Y.}\ \bibnamefont
  {Kitaev}},\ }\href {\doibase 10.1070/1063-7869/44/10s/s29} {\bibfield
  {journal} {\bibinfo  {journal} {Phys. Uspekhi}\ }\textbf {\bibinfo {volume}
  {44}},\ \bibinfo {pages} {131} (\bibinfo {year} {2001})},\ \Eprint
  {http://arxiv.org/abs/0010440} {arXiv:0010440 [cond-mat]} \BibitemShut
  {NoStop}%
\bibitem [{\citenamefont {Kwon}\ \emph {et~al.}(2004)\citenamefont {Kwon},
  \citenamefont {Sengupta},\ and\ \citenamefont {Yakovenko}}]{Kwon2004}%
  \BibitemOpen
  \bibfield  {author} {\bibinfo {author} {\bibfnamefont {H.~J.}\ \bibnamefont
  {Kwon}}, \bibinfo {author} {\bibfnamefont {K.}~\bibnamefont {Sengupta}}, \
  and\ \bibinfo {author} {\bibfnamefont {V.~M.}\ \bibnamefont {Yakovenko}},\
  }\href {\doibase 10.1140/epjb/e2004-00066-4} {\bibfield  {journal} {\bibinfo
  {journal} {Eur. Phys. J. B}\ }\textbf {\bibinfo {volume} {37}},\ \bibinfo
  {pages} {349} (\bibinfo {year} {2004})},\ \Eprint
  {http://arxiv.org/abs/0210148} {arXiv:0210148 [cond-mat]} \BibitemShut
  {NoStop}%
\bibitem [{\citenamefont {Deacon}\ \emph {et~al.}(2017)\citenamefont {Deacon},
  \citenamefont {Wiedenmann}, \citenamefont {Bocquillon}, \citenamefont
  {Dom{\'{i}}nguez}, \citenamefont {Klapwijk}, \citenamefont {Leubner},
  \citenamefont {Br{\"{u}}ne}, \citenamefont {Hankiewicz}, \citenamefont
  {Tarucha}, \citenamefont {Ishibashi}, \citenamefont {Buhmann},\ and\
  \citenamefont {Molenkamp}}]{Deacon2017JosephsonJunctions}%
  \BibitemOpen
  \bibfield  {author} {\bibinfo {author} {\bibfnamefont {R.~S.}\ \bibnamefont
  {Deacon}}, \bibinfo {author} {\bibfnamefont {J.}~\bibnamefont {Wiedenmann}},
  \bibinfo {author} {\bibfnamefont {E.}~\bibnamefont {Bocquillon}}, \bibinfo
  {author} {\bibfnamefont {F.}~\bibnamefont {Dom{\'{i}}nguez}}, \bibinfo
  {author} {\bibfnamefont {T.~M.}\ \bibnamefont {Klapwijk}}, \bibinfo {author}
  {\bibfnamefont {P.}~\bibnamefont {Leubner}}, \bibinfo {author} {\bibfnamefont
  {C.}~\bibnamefont {Br{\"{u}}ne}}, \bibinfo {author} {\bibfnamefont {E.~M.}\
  \bibnamefont {Hankiewicz}}, \bibinfo {author} {\bibfnamefont
  {S.}~\bibnamefont {Tarucha}}, \bibinfo {author} {\bibfnamefont
  {K.}~\bibnamefont {Ishibashi}}, \bibinfo {author} {\bibfnamefont
  {H.}~\bibnamefont {Buhmann}}, \ and\ \bibinfo {author} {\bibfnamefont
  {L.~W.}\ \bibnamefont {Molenkamp}},\ }\href {\doibase
  10.1103/PhysRevX.7.021011} {\bibfield  {journal} {\bibinfo  {journal} {Phys.
  Rev. X}\ }\textbf {\bibinfo {volume} {7}},\ \bibinfo {pages} {21011}
  (\bibinfo {year} {2017})}\BibitemShut {NoStop}%
\bibitem [{\citenamefont {Laroche}\ \emph {et~al.}(2019)\citenamefont
  {Laroche}, \citenamefont {Bouman}, \citenamefont {van Woerkom}, \citenamefont
  {Proutski}, \citenamefont {Murthy}, \citenamefont {Pikulin}, \citenamefont
  {Nayak}, \citenamefont {van Gulik}, \citenamefont {Nyg{\aa}rd}, \citenamefont
  {Krogstrup}, \citenamefont {Kouwenhoven},\ and\ \citenamefont
  {Geresdi}}]{Laroche2019ObservationNanowires}%
  \BibitemOpen
  \bibfield  {author} {\bibinfo {author} {\bibfnamefont {D.}~\bibnamefont
  {Laroche}}, \bibinfo {author} {\bibfnamefont {D.}~\bibnamefont {Bouman}},
  \bibinfo {author} {\bibfnamefont {D.~J.}\ \bibnamefont {van Woerkom}},
  \bibinfo {author} {\bibfnamefont {A.}~\bibnamefont {Proutski}}, \bibinfo
  {author} {\bibfnamefont {C.}~\bibnamefont {Murthy}}, \bibinfo {author}
  {\bibfnamefont {D.~I.}\ \bibnamefont {Pikulin}}, \bibinfo {author}
  {\bibfnamefont {C.}~\bibnamefont {Nayak}}, \bibinfo {author} {\bibfnamefont
  {R.~J.~J.}\ \bibnamefont {van Gulik}}, \bibinfo {author} {\bibfnamefont
  {J.}~\bibnamefont {Nyg{\aa}rd}}, \bibinfo {author} {\bibfnamefont
  {P.}~\bibnamefont {Krogstrup}}, \bibinfo {author} {\bibfnamefont {L.~P.}\
  \bibnamefont {Kouwenhoven}}, \ and\ \bibinfo {author} {\bibfnamefont
  {A.}~\bibnamefont {Geresdi}},\ }\href {\doibase 10.1038/s41467-018-08161-2}
  {\bibfield  {journal} {\bibinfo  {journal} {Nat. Commun.}\ }\textbf {\bibinfo
  {volume} {10}},\ \bibinfo {pages} {245} (\bibinfo {year} {2019})}\BibitemShut
  {NoStop}%
\bibitem [{\citenamefont {Fu}\ and\ \citenamefont {Kane}(2009)}]{Fu2009}%
  \BibitemOpen
  \bibfield  {author} {\bibinfo {author} {\bibfnamefont {L.}~\bibnamefont
  {Fu}}\ and\ \bibinfo {author} {\bibfnamefont {C.~L.}\ \bibnamefont {Kane}},\
  }\href {\doibase 10.1103/PhysRevB.79.161408} {\bibfield  {journal} {\bibinfo
  {journal} {Phys. Rev. B}\ }\textbf {\bibinfo {volume} {79}},\ \bibinfo
  {pages} {161408(R)} (\bibinfo {year} {2009})},\ \Eprint
  {http://arxiv.org/abs/0804.4469} {arXiv:0804.4469} \BibitemShut {NoStop}%
\bibitem [{\citenamefont {Badiane}\ \emph {et~al.}(2011)\citenamefont
  {Badiane}, \citenamefont {Houzet},\ and\ \citenamefont
  {Meyer}}]{Badiane2011}%
  \BibitemOpen
  \bibfield  {author} {\bibinfo {author} {\bibfnamefont {D.~M.}\ \bibnamefont
  {Badiane}}, \bibinfo {author} {\bibfnamefont {M.}~\bibnamefont {Houzet}}, \
  and\ \bibinfo {author} {\bibfnamefont {J.~S.}\ \bibnamefont {Meyer}},\ }\href
  {\doibase 10.1103/PhysRevLett.107.177002} {\bibfield  {journal} {\bibinfo
  {journal} {Phys. Rev. Lett.}\ }\textbf {\bibinfo {volume} {107}},\ \bibinfo
  {pages} {177002} (\bibinfo {year} {2011})},\ \Eprint
  {http://arxiv.org/abs/1108.3870} {arXiv:1108.3870} \BibitemShut {NoStop}%
\bibitem [{\citenamefont {Pikulin}\ and\ \citenamefont
  {Nazarov}(2012)}]{Pikulin2012}%
  \BibitemOpen
  \bibfield  {author} {\bibinfo {author} {\bibfnamefont {D.~I.}\ \bibnamefont
  {Pikulin}}\ and\ \bibinfo {author} {\bibfnamefont {Y.~V.}\ \bibnamefont
  {Nazarov}},\ }\href {\doibase 10.1103/PhysRevB.86.140504} {\bibfield
  {journal} {\bibinfo  {journal} {Phys. Rev. B}\ }\textbf {\bibinfo {volume}
  {86}},\ \bibinfo {pages} {140504(R)} (\bibinfo {year} {2012})}\BibitemShut
  {NoStop}%
\bibitem [{\citenamefont {San-Jose}\ \emph {et~al.}(2012)\citenamefont
  {San-Jose}, \citenamefont {Prada},\ and\ \citenamefont
  {Aguado}}]{San-Jose2012}%
  \BibitemOpen
  \bibfield  {author} {\bibinfo {author} {\bibfnamefont {P.}~\bibnamefont
  {San-Jose}}, \bibinfo {author} {\bibfnamefont {E.}~\bibnamefont {Prada}}, \
  and\ \bibinfo {author} {\bibfnamefont {R.}~\bibnamefont {Aguado}},\ }\href
  {\doibase 10.1103/PhysRevLett.108.257001} {\bibfield  {journal} {\bibinfo
  {journal} {Phys. Rev. Lett.}\ }\textbf {\bibinfo {volume} {108}},\ \bibinfo
  {pages} {257001} (\bibinfo {year} {2012})},\ \Eprint
  {http://arxiv.org/abs/1112.5983} {arXiv:1112.5983} \BibitemShut {NoStop}%
\bibitem [{\citenamefont {Mart{\'{i}}n-Rodero}\ and\ \citenamefont {{Levy
  Yeyati}}(2011)}]{Martin-Rodero2011}%
  \BibitemOpen
  \bibfield  {author} {\bibinfo {author} {\bibfnamefont {A.}~\bibnamefont
  {Mart{\'{i}}n-Rodero}}\ and\ \bibinfo {author} {\bibfnamefont
  {A.}~\bibnamefont {{Levy Yeyati}}},\ }\href {\doibase
  10.1080/00018732.2011.624266} {\bibfield  {journal} {\bibinfo  {journal}
  {Adv. Physique}\ }\textbf {\bibinfo {volume} {60}},\ \bibinfo {pages} {899}
  (\bibinfo {year} {2011})},\ \Eprint {http://arxiv.org/abs/1111.4939}
  {arXiv:1111.4939} \BibitemShut {NoStop}%
\bibitem [{\citenamefont {Glazman}\ and\ \citenamefont
  {Matveev}(1989)}]{Glazman1989ResonantBarrier}%
  \BibitemOpen
  \bibfield  {author} {\bibinfo {author} {\bibfnamefont {L.~I.}\ \bibnamefont
  {Glazman}}\ and\ \bibinfo {author} {\bibfnamefont {A.}~\bibnamefont
  {Matveev}},\ }\bibfield  {booktitle} {\emph {\bibinfo {booktitle} {J. Exp.
  Theor. Phys.}},\ }\href
  {https://ui.adsabs.harvard.edu/abs/1989ZhPmR..49..570G/abstract} {\bibfield
  {journal} {\bibinfo  {journal} {JETP Lett.}\ }\textbf {\bibinfo {volume}
  {49}},\ \bibinfo {pages} {659} (\bibinfo {year} {1989})}\BibitemShut
  {NoStop}%
\bibitem [{\citenamefont {Vecino}\ \emph {et~al.}(2003)\citenamefont {Vecino},
  \citenamefont {Mart{\'{i}}n-Rodero},\ and\ \citenamefont {{Levy
  Yeyati}}}]{Vecino2003}%
  \BibitemOpen
  \bibfield  {author} {\bibinfo {author} {\bibfnamefont {E.}~\bibnamefont
  {Vecino}}, \bibinfo {author} {\bibfnamefont {A.}~\bibnamefont
  {Mart{\'{i}}n-Rodero}}, \ and\ \bibinfo {author} {\bibfnamefont
  {A.}~\bibnamefont {{Levy Yeyati}}},\ }\href {\doibase
  10.1103/PhysRevB.68.035105} {\bibfield  {journal} {\bibinfo  {journal} {Phys.
  Rev. B}\ }\textbf {\bibinfo {volume} {68}},\ \bibinfo {pages} {035105}
  (\bibinfo {year} {2003})},\ \Eprint {http://arxiv.org/abs/0212375}
  {arXiv:0212375 [cond-mat]} \BibitemShut {NoStop}%
\bibitem [{\citenamefont {Zazunov}\ \emph {et~al.}(2018)\citenamefont
  {Zazunov}, \citenamefont {Plugge},\ and\ \citenamefont
  {Egger}}]{Zazunov2018}%
  \BibitemOpen
  \bibfield  {author} {\bibinfo {author} {\bibfnamefont {A.}~\bibnamefont
  {Zazunov}}, \bibinfo {author} {\bibfnamefont {S.}~\bibnamefont {Plugge}}, \
  and\ \bibinfo {author} {\bibfnamefont {R.}~\bibnamefont {Egger}},\ }\href
  {\doibase 10.1103/PhysRevLett.121.207701} {\bibfield  {journal} {\bibinfo
  {journal} {Phys. Rev. Lett.}\ }\textbf {\bibinfo {volume} {121}},\ \bibinfo
  {pages} {207701} (\bibinfo {year} {2018})},\ \Eprint
  {http://arxiv.org/abs/1809.06892} {arXiv:1809.06892} \BibitemShut {NoStop}%
\bibitem [{\citenamefont {Matsuura}(1977)}]{Matsuura1977}%
  \BibitemOpen
  \bibfield  {author} {\bibinfo {author} {\bibfnamefont {T.}~\bibnamefont
  {Matsuura}},\ }\href {\doibase 10.1143/ptp.57.1823} {\bibfield  {journal}
  {\bibinfo  {journal} {Prog. Theor. Phys.}\ }\textbf {\bibinfo {volume}
  {57}},\ \bibinfo {pages} {1823} (\bibinfo {year} {1977})}\BibitemShut
  {NoStop}%
\bibitem [{\citenamefont {Beenakker}\ and\ \citenamefont {van
  Houten}(1992)}]{Beenakker1991}%
  \BibitemOpen
  \bibfield  {author} {\bibinfo {author} {\bibfnamefont {C.~W.~J.}\
  \bibnamefont {Beenakker}}\ and\ \bibinfo {author} {\bibfnamefont
  {H.}~\bibnamefont {van Houten}},\ }in\ \href {\doibase
  10.1007/978-3-642-77274-0_20} {\emph {\bibinfo {booktitle} {Single-Electron
  Tunneling and Mesoscopic Devices. Springer Series in Electronics and
  Photonics}}},\ \bibinfo {editor} {edited by\ \bibinfo {editor} {\bibfnamefont
  {H.}~\bibnamefont {Koch}}\ and\ \bibinfo {editor} {\bibfnamefont
  {H.}~\bibnamefont {L{\"{u}}bbig}}}\ (\bibinfo  {publisher} {Springer},\
  \bibinfo {address} {Berlin},\ \bibinfo {year} {1992})\ pp.\ \bibinfo {pages}
  {175--179},\ \Eprint {http://arxiv.org/abs/0111505} {arXiv:0111505
  [cond-mat]} \BibitemShut {NoStop}%
\bibitem [{\citenamefont {Averin}\ and\ \citenamefont
  {Bardas}(1995)}]{Averin1995}%
  \BibitemOpen
  \bibfield  {author} {\bibinfo {author} {\bibfnamefont {D.}~\bibnamefont
  {Averin}}\ and\ \bibinfo {author} {\bibfnamefont {A.}~\bibnamefont
  {Bardas}},\ }\href {\doibase 10.1103/PhysRevLett.75.1831} {\bibfield
  {journal} {\bibinfo  {journal} {Phys. Rev. Lett.}\ }\textbf {\bibinfo
  {volume} {75}},\ \bibinfo {pages} {1831} (\bibinfo {year}
  {1995})}\BibitemShut {NoStop}%
\bibitem [{\citenamefont {Houzet}\ \emph {et~al.}(2013)\citenamefont {Houzet},
  \citenamefont {Meyer}, \citenamefont {Badiane},\ and\ \citenamefont
  {Glazman}}]{Houzet2013}%
  \BibitemOpen
  \bibfield  {author} {\bibinfo {author} {\bibfnamefont {M.}~\bibnamefont
  {Houzet}}, \bibinfo {author} {\bibfnamefont {J.~S.}\ \bibnamefont {Meyer}},
  \bibinfo {author} {\bibfnamefont {D.~M.}\ \bibnamefont {Badiane}}, \ and\
  \bibinfo {author} {\bibfnamefont {L.~I.}\ \bibnamefont {Glazman}},\ }\href
  {\doibase 10.1103/PhysRevLett.111.046401} {\bibfield  {journal} {\bibinfo
  {journal} {Phys. Rev. Lett.}\ }\textbf {\bibinfo {volume} {111}},\ \bibinfo
  {pages} {046401} (\bibinfo {year} {2013})},\ \Eprint
  {http://arxiv.org/abs/1303.4909} {arXiv:1303.4909} \BibitemShut {NoStop}%
\bibitem [{\citenamefont {{Levy Yeyati}}\ \emph {et~al.}(2003)\citenamefont
  {{Levy Yeyati}}, \citenamefont {Mart{\'{i}}n-Rodero},\ and\ \citenamefont
  {Vecino}}]{Yeyati2003}%
  \BibitemOpen
  \bibfield  {author} {\bibinfo {author} {\bibfnamefont {A.}~\bibnamefont
  {{Levy Yeyati}}}, \bibinfo {author} {\bibfnamefont {A.}~\bibnamefont
  {Mart{\'{i}}n-Rodero}}, \ and\ \bibinfo {author} {\bibfnamefont
  {E.}~\bibnamefont {Vecino}},\ }\href {\doibase 10.1103/PhysRevLett.91.266802}
  {\bibfield  {journal} {\bibinfo  {journal} {Phys. Rev. Lett.}\ }\textbf
  {\bibinfo {volume} {91}},\ \bibinfo {pages} {266802} (\bibinfo {year}
  {2003})}\BibitemShut {NoStop}%
\bibitem [{\citenamefont {Averin}\ and\ \citenamefont
  {Imam}(1996)}]{Averin1996}%
  \BibitemOpen
  \bibfield  {author} {\bibinfo {author} {\bibfnamefont {D.}~\bibnamefont
  {Averin}}\ and\ \bibinfo {author} {\bibfnamefont {H.~T.}\ \bibnamefont
  {Imam}},\ }\href {\doibase 10.1103/PhysRevLett.76.3814} {\bibfield  {journal}
  {\bibinfo  {journal} {Phys. Rev. Lett.}\ }\textbf {\bibinfo {volume} {76}},\
  \bibinfo {pages} {3814} (\bibinfo {year} {1996})}\BibitemShut {NoStop}%
\bibitem [{sup()}]{supplemental}%
  \BibitemOpen
  \href@noop {} {}\bibinfo {note} {{See supplemental material for the
  derivation of equation (14) and (15) and an alternative description of the
  junction at $p_\pi = 1$, which clarifies the relation between our model and
  topological junctions.}}\BibitemShut {Stop}%
\bibitem [{\citenamefont {Naveh}\ and\ \citenamefont
  {Averin}(1999)}]{Naveh1999}%
  \BibitemOpen
  \bibfield  {author} {\bibinfo {author} {\bibfnamefont {Y.}~\bibnamefont
  {Naveh}}\ and\ \bibinfo {author} {\bibfnamefont {D.~V.}\ \bibnamefont
  {Averin}},\ }\href {\doibase 10.1103/PhysRevLett.82.4090} {\bibfield
  {journal} {\bibinfo  {journal} {Phys. Rev. Lett.}\ }\textbf {\bibinfo
  {volume} {82}},\ \bibinfo {pages} {4090} (\bibinfo {year} {1999})},\ \Eprint
  {http://arxiv.org/abs/9902202} {arXiv:9902202 [cond-mat]} \BibitemShut
  {NoStop}%
\bibitem [{\citenamefont {Hata}\ \emph {et~al.}(2018)\citenamefont {Hata},
  \citenamefont {Delagrange}, \citenamefont {Arakawa}, \citenamefont {Lee},
  \citenamefont {Deblock}, \citenamefont {Bouchiat}, \citenamefont
  {Kobayashi},\ and\ \citenamefont {Ferrier}}]{Hata2018}%
  \BibitemOpen
  \bibfield  {author} {\bibinfo {author} {\bibfnamefont {T.}~\bibnamefont
  {Hata}}, \bibinfo {author} {\bibfnamefont {R.}~\bibnamefont {Delagrange}},
  \bibinfo {author} {\bibfnamefont {T.}~\bibnamefont {Arakawa}}, \bibinfo
  {author} {\bibfnamefont {S.}~\bibnamefont {Lee}}, \bibinfo {author}
  {\bibfnamefont {R.}~\bibnamefont {Deblock}}, \bibinfo {author} {\bibfnamefont
  {H.}~\bibnamefont {Bouchiat}}, \bibinfo {author} {\bibfnamefont
  {K.}~\bibnamefont {Kobayashi}}, \ and\ \bibinfo {author} {\bibfnamefont
  {M.}~\bibnamefont {Ferrier}},\ }\href {\doibase
  10.1103/PhysRevLett.121.247703} {\bibfield  {journal} {\bibinfo  {journal}
  {Phys. Rev. Lett.}\ }\textbf {\bibinfo {volume} {121}},\ \bibinfo {pages}
  {247703} (\bibinfo {year} {2018})},\ \Eprint
  {http://arxiv.org/abs/1805.07853} {arXiv:1805.07853} \BibitemShut {NoStop}%
\end{thebibliography}%
	
\end{document}


\definecolor{rvwvcq}{rgb}{0.08235294117647059,0.396078431372549,0.7529411764705882}

\begin{center}
{\large\bf SUPPLEMENTAL MATERIAL: 

\smallskip

Josephson radiation in a superconductor-quantum dot-superconductor junction}

\medskip

Baptiste~Lamic, Julia~S.~Meyer, and Manuel~Houzet

{\it Univ.~Grenoble Alpes, CEA, IRIG-Pheliqs, F-38000 Grenoble, France}

\smallskip

\today
\end{center}

In section~\ref{SM-1} of this supplemental material, we provide technical details on the derivation of our results for the current noise spectrum, Eqs.~(14) and (15) of the main text. In section~\ref{SM-2}, we propose an alternative description of the junction at Landau-Zener probability $p_\pi = 1$ that clarifies the relation between our model and topological junctions. 

\section{Evaluation of the current noise spectrum}\label{SM-1}

Combining Eqs.~(11), (12), and (13) of the main text, one obtains the following expression for the noise:
\begin{equation}
   { S(t_1,t_2) =} {\cal \bm{I}}(t_1) {\cal \bm{I}}(t_2) \sum_{\alpha=1}^{3}v_0^{T}{\cal \bm{\hat{I}}} {\cal \bm{L}}_\pi^{k_1} u_\alpha\lambda_\alpha^{m_1-m_2}v_\alpha^T {\cal \bm{L}}_\pi^{-k_2} {\cal\bm{\hat I}}{\cal L}_\pi^{k_2}u_0,
    \label{eq:full_cor}
\end{equation}
for $(2 m_i+k_i)\pi<\omega_Jt_i + \varphi_0<(2 m_i+k_i+1)\pi$. 
Here ${\cal \bm{\hat{I}}}$ is the current operator, which in the basis of  Eqs.~(4), (5) takes the form
\begin{equation}
{\cal \bm{\hat{I}}} = 
\left(\begin{array}{cccc}
0&0&0&0
\\0&-1 &0&0
\\0&0&1&0
\\0&0&0&0
\end{array}\right).
\end{equation}
To evaluate the noise, the left and right eigenvectors of the stochastic matric ${\cal L}$ are needed. The right eigenvector $u_0$, describing the stationary state is given in the main text, whereas the others read as follows:
\begin{equation}
u_1 = \left(\begin{array}{c}
1\\
0\\
0\\
-1
\end{array}\right),
\qquad
u_2 = \left(\begin{array}{c}
1\\
-1\\
-1\\
1
\end{array}\right),
\qquad
u_3 = \left(\begin{array}{c}
p_0(1-2p_\pi)\\
-2p_\pi(1-p_0)\\
-2p_0+2p_\pi(1+p_0)\\
p_0(1-2p_\pi)
\end{array}\right).
\label{eq:u}
\end{equation}
One notes that the eigenvectors $u_1$ and $u_2$ do not carry any current, $v_0^T{\cal \bm{\hat I}} u_{1,2} =0$. Furthermore, $u_1$ and $u_2$ are eigenvectors not only of ${\cal L}$ but also of ${\cal L}_\pi$, namely ${\cal L}_\pi u_{1,2}=u_{1,2}$, which further yields $v_0^T{\cal \bm{\hat I}} {\cal L}_\pi u_{1,2} = 0$. As a consequence, the expression for the noise reduces to 
\begin{equation}
 S(t_1,t_2) =  
{\cal \bm{I}}(t_1) {\cal \bm{I}}(t_2) \lambda_3^{m_1-m_2} \left(v_0^T {\cal \bm{\hat{I}}} {\cal \bm{L}}_\pi^{k_1}u_3\right) \left(v_3^T {\cal \bm{L}}_\pi^{-k_2} {\cal\bm{\hat I}} \mathcal{L}_\pi^{k_2} u_0\right).
\label{eq:full_cor_sim}
\end{equation}
Only the eigenvalue $\lambda_3$ contributes to the current fluctuation. Thus,  there is a unique time scale for the decay of correlations. In addition to $v_0$ given in the main text, the only other left eigenvector needed is $v_3$, given as
\begin{equation}
v_3= -\frac{ 1}{2 \left(
p_0-2 p_{\pi }\right) \left(1-\lambda_3\right)}
\left(
\begin{array}{c}
p_0\\
2p_0 +2p_\pi(1-p_0)\\
-2p_\pi(1-p_0) \\
 p_0
\end{array}\right).
\label{eq:v}
\end{equation}
With the expressions provided in Eqs.~\eqref{eq:u} and \eqref{eq:v}, as well as Eqs.~(5) and (7) of the main text, we can evaluate the matrix elements contributing to the noise as given in Eq.~\eqref{eq:full_cor_sim}. In particular,
\begin{subequations}
\label{eq:vIu}
	\begin{eqnarray}
	v_0^{T}{\cal \bm{\hat{I}}} u_3 &=& 2(2p_\pi-p_0),\\
	v_3^T {\cal\bm{\hat I}} u_0  &=&	 \frac{\lambda_3p_\pi[(1-p_0)(2p_\pi-p_0)-\mathcal{N}]}{{\cal N}(p_0-2p_\pi) (1-2p_\pi)(1-\lambda_3)} ,\\
	 	v_3^T {\cal \bm{L}}_\pi^{-1} {\cal\bm{\hat I}} \mathcal{L}_\pi u_0 &=& \frac{p_\pi(2p_\pi-p_0-{\cal N})}{{\cal N}(p_0-2p_\pi) (1-2p_\pi)(1-\lambda_3)} ,
			\end{eqnarray}
\end{subequations}
and $v_0^{T}{\cal \bm{\hat{I}}} {\cal \bm{L}}_\pi u_3 = \left(1-2 p_{\pi }\right)  v_0^{T}{\cal \bm{\hat{I}}} u_3$.

We can now turn to the computation of the noise spectrum $S(\omega)$. As a first step, we rewrite Eq.~(10) of the main text as
\begin{equation}
S(\omega) =\lim_{T\rightarrow \infty}\frac{2}{T} \int_0^\infty  d\tau \; \cos(\omega\tau)\int_{0}^{T} dt \;S(t+\tau,t).
\end{equation}
The function $S(t+\tau,t)$ is  periodic in $t$ and quasi-periodic in $\tau$ with period $T=2\pi/\omega_J$: 
\begin{eqnarray}
	 S(t+ 2\pi/\omega_J+\tau  ,t + 2\pi/\omega_J) &=&S(t+\tau,t),  \label{eq:periodic}\\
	 S(t+\tau+\frac{2\pi}{\omega_J},t)&=&\lambda_3 S(t+\tau,t).   \label{eq:quasiperiodic}
\end{eqnarray}
Using the periodicity in $t$, one may rewrite the time average as
\begin{equation}
\label{eq:periodic}
\lim_{T\rightarrow \infty} \frac{1}{T}\int_{0}^{T} dt \;S(t+\tau,t) = \frac{\omega_J}{2\pi}\int_{0}^{2\pi/\omega_J} dt \;S(t+\tau,t).
\end{equation}
Furthermore, using the quasi-periodicity in $\tau$, we may rewrite the Fourier transform as 
\begin{equation}
\label{eq:quasiperiodic}
\int_{0}^{\infty} d\tau \;\cos(\omega \tau) S(t+\tau,t) = \mathcal{R}\left\{\sum_{n=0}^{\infty}\left(\lambda_3 e^{2i\pi\omega/\omega_J}\right)^n\int_{0}^{2\pi/\omega_J} d\tau\; e^{i\omega \tau} S(t+\tau,t)\right\}.
\end{equation}
Combining Eqs.~\eqref{eq:periodic} and \eqref{eq:quasiperiodic} and performing the sum over $n$, one obtains
\begin{equation}
S(\omega) = \frac{\omega_J}{\pi}\mathcal{R}\left\{\frac{1}{1-\lambda_3 e^{\frac{2i\pi\omega}{\omega_J}}}\int_{0}^{2\pi/\omega_J} d\tau \int_{0}^{2\pi/\omega_J} dt\; e^{i\omega\tau} S(t+\tau,t)\right\}.
\end{equation}
 With  the changes of variables $\omega_Jt +\varphi_0 \rightarrow \varphi$ and  $\omega_J\tau \rightarrow \delta$, we can write
 \begin{equation}
 	\int_{0}^{2\pi/\omega_J} d\tau \int_{0}^{2\pi/\omega_J}\, dt\, e^{i\omega\tau} S(t+\tau,t) = \frac{1}{\omega_J^2} \int_{0}^{2\pi} d\delta \int_{0}^{2\pi}\, d\varphi\, e^{i\tilde\omega\tau} S(\varphi+\delta,\varphi),
 \end{equation}
 where $\tilde\omega = \omega/\omega_J$.
 To perform the remaining double integral, one notices that $S(t,t+\tau)$ possesses discontinuities at
 $\delta + \varphi = n\pi$ and $\varphi = n\pi$. At these phases, the values of the indices $m_1-m_2$, $k_1$, and $k_2$ change in the expression for the noise, Eq.~\eqref{eq:full_cor_sim}. Thus, the integration domain can  be split into different segments.
  \begin{figure}[h]
 	\centering
		\includegraphics[width=0.49\linewidth]{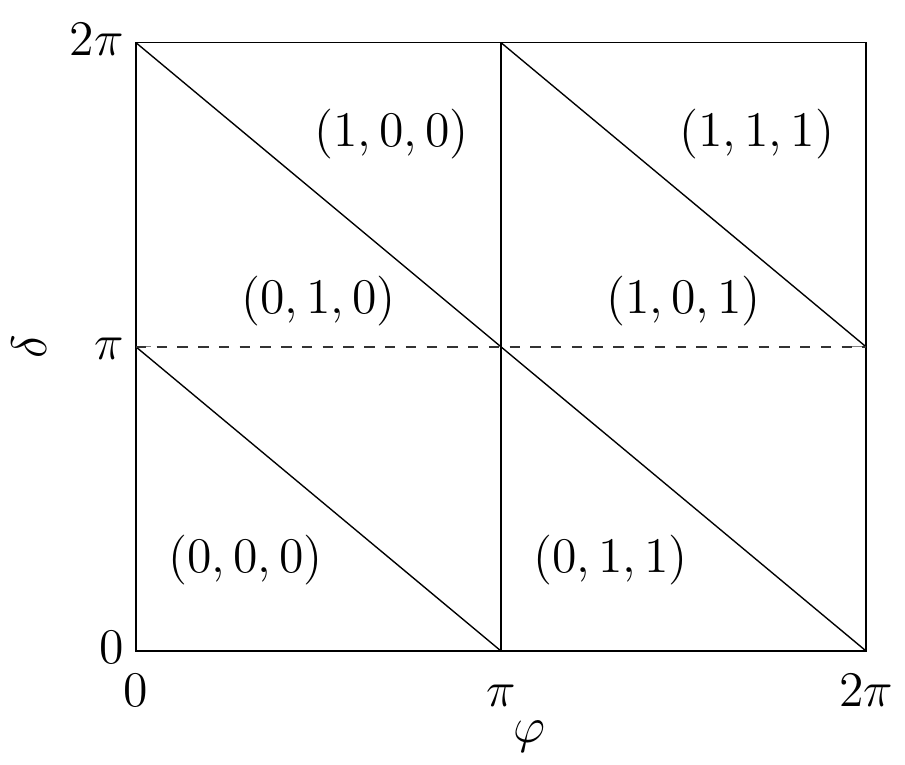}
 	\caption{
 		\label{fig:indices}
 		Map of the indices $(m_1-m_2,k_1,k_2)$ determining $S(\varphi+\delta,\varphi)$ according to Eq.~\eqref{eq:full_cor_sim} as a function of $\varphi$ and $\delta$. Discontinuities in $S(\varphi + \delta,\varphi)$ appear at the boundaries of the different domains delimited by full lines. The integration scheme introduces additional segments depending on the value of $\delta$, separated by the dashed line.
 	}
 \end{figure}
 Performing the integration over $\varphi$ in the eight segments shown in Fig.~\ref{fig:indices}, we end up with a piecewise continuous function of $\delta$. 
Integrating over $\delta$ and rearranging the terms, we obtain Eq.~(14) of the main text with coefficients
\newcommand{\vILu}{v_0^{T}{\cal \bm{\hat{I}}} u_3}
\newcommand{\vLILuZ}{v_3^T {\cal\bm{\hat I}} u_0}
\newcommand{\vLILuO}{v_3^T {\cal \bm{L}}_\pi^{-1} {\cal\bm{\hat I}} \mathcal{L}_\pi u_0}
\begin{dgroup}
\begin{dmath}
A = 2 \left(1+\lambda _3\right) \vILu \left\{\left[-2\lambda_3  +(1-2p_\pi)(1-\lambda _3)\right] \vLILuO+\left[1-\lambda_3+2(1-2p_\pi)\right] \vLILuZ\right\},
\end{dmath}
\begin{dmath}
B= 8 \vILu \left[\lambda _3^2 \vLILuO-\left(1-2 p_{\pi }\right) \vLILuZ\right],
\end{dmath}
\begin{dmath}
C= 32 \lambda _3 \vILu \left[\vLILuO-\left(1-2 p_{\pi }\right) \vLILuZ\right].
\end{dmath}
\end{dgroup}
Using Eqs.~\eqref{eq:vIu}, this yields the expressions for the coefficients given in Eqs.~(15) of the main text.

\section{Equivalent model at $p_\pi = 1$}\label{SM-2}

When the Landau-Zener probablity is $p_\pi=1$, stochastic processes happen only at phases $\varphi=2m\pi$. By contrast, at $\varphi=(2m+1)\pi$, a deterministic exchange between the states $\ket{\pm}$ takes place. In this case, a more natural basis of states $\ket{a},\ket{b}$ is given as
\begin{itemize}
	\item $\ket{a} = \ket{+}$ and $\ket{b} = \ket{-}$ for $(4m-1)\pi<\varphi< (4m+1)\pi$,
	\item $\ket{a} = \ket{-}$ and $\ket{b} = \ket{+}$ for $(4m+1)\pi<\varphi< 4(m+3)\pi$,
\end{itemize} 
which absorbs the effect of the deterministic Landau-Zener transitions due to ${\cal L}_\pi$ at $p_\pi=1$. 

Then the probabilities $\tilde{\bm{P}}(\varphi)=\left(P_{1_a 1_b}(\varphi),P_{1_a 0_b}(\varphi),P_{0_a 1_b}(\varphi),P_{0_a 0_b}(\varphi)\right)^T$ are constant over the phase intervals $2m\pi < \varphi < 2(m+1)\pi$ and evolve according to 
\begin{subequations}
	\begin{eqnarray}
	\tilde{\bm{P}}(4m\pi^+)&=&\mathcal{L}_0\tilde{\bm{P}}(4m\pi ^-) ,\\
	\tilde{\bm{P}}((4m+2)\pi^+)&=& \mathcal{L}_{2\pi}\tilde{\bm{P}}((4m+2)\pi ^-).
	\end{eqnarray} 
\end{subequations}
Here the transition matrices ${\cal L}_\varphi$ can be decomposed into a product of Markov matrices acting on distinct ABS branches, ${\cal L}_\varphi={\cal L}_\varphi^{(a)}\otimes{\cal L}_\varphi^{(b)}$. In particular,
	\begin{eqnarray}
\mathcal{L}_0 = \mathcal{L_-}\otimes\mathcal{L_+},
 \qquad
\mathcal{L}_{2\pi} = \mathcal{L_+}\otimes\mathcal{L_-},
\end{eqnarray}
 where the new transition matrices in the basis $\left\{\ket{1}_x,\ket{0}_x\right\}$ with $x=a,b$ take the form  
\begin{eqnarray}
	\mathcal{L_-} = \left(\begin{array}{cc}
	1 & p_0 \\
	0 & 1-p_0
	\end{array}\right),
	\qquad
	\mathcal{L_+} = \left(\begin{array}{cc}
	1-p_0 & 0 \\
	p_0 & 1
	\end{array}\right).
\end{eqnarray} 
\begin{figure}[!h]
		\includegraphics[width=0.8\linewidth]{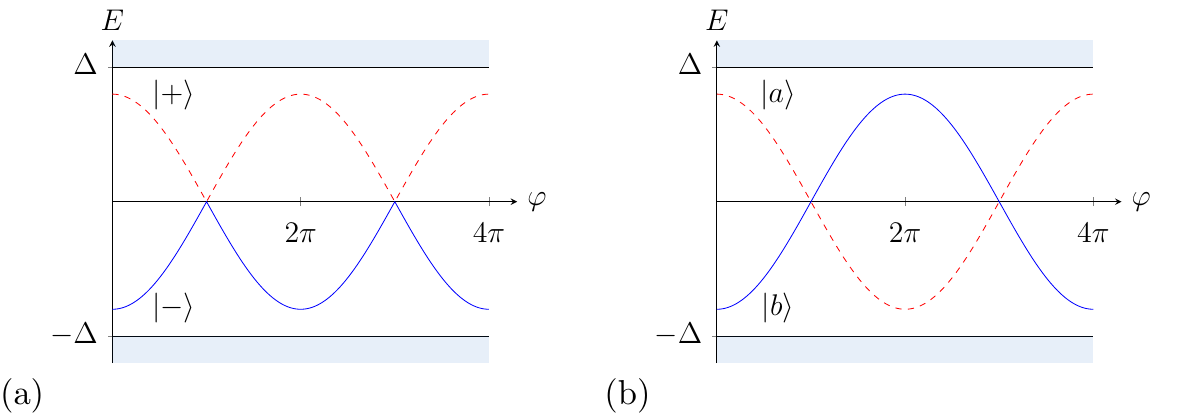}
	\caption{ (a) Representation of the Andreev Bound states in the initial basis $|\pm\rangle$. (b) Representation of the Andreev Bound states in  the basis $|a/b\rangle$. The deterministic Landau-Zener process at $\varphi = (2m + 1)\pi$ have been absorbed by the change of basis.}
\end{figure}

Thus the occupations of the two states $\ket{a}$ and $\ket{b}$ switch independently. The dynamics of each state is governed by the same equations as the dynamics of the $4\pi$-periodic Andreev bound state appearing in a topological Josephson junction~[S1]. The two states carry a current $I_{a/b}(\varphi)=\pm I_J\sin(\varphi/2)$. Furthermore, the current operator ${\cal \bm{\hat{I}}}$ is also separable,
\begin{equation}
	{\cal \bm{\hat{I}}} = \frac12\left(\sigma_z \otimes\mathbbm{1} - \mathbbm{1} \otimes\sigma_z\right).
\end{equation} 
As a consequence, at $p_\pi = 1$, the system can be described as two independent topological junctions in parallel. Thus, both the average current as well as the noise are doubled compared to the values for the topological junction reported in~[S1]. As a consequence, one obtains the same Fano factor, $F=1/2$, as well as a peak in the noise spectrum at $\omega=\omega_J/2$, corresponding to a fractional Josephson effect.